\PassOptionsToPackage{unicode}{hyperref}
\PassOptionsToPackage{hyphens}{url}
\PassOptionsToPackage{dvipsnames,svgnames,x11names}{xcolor}
\documentclass[
  11pt,
  a4paper,
]{article}
\usepackage{amsmath,amssymb}
\usepackage{lmodern}
\usepackage{iftex}
\ifPDFTeX
  \usepackage[T1]{fontenc}
  \usepackage[utf8]{inputenc}
  \usepackage{textcomp} % provide euro and other symbols
\else % if luatex or xetex
  \usepackage{unicode-math}
  \defaultfontfeatures{Scale=MatchLowercase}
  \defaultfontfeatures[\rmfamily]{Ligatures=TeX,Scale=1}
\fi
\IfFileExists{upquote.sty}{\usepackage{upquote}}{}
\IfFileExists{microtype.sty}{% use microtype if available
  \usepackage[]{microtype}
  \UseMicrotypeSet[protrusion]{basicmath} % disable protrusion for tt fonts
}{}
\makeatletter
\@ifundefined{KOMAClassName}{% if non-KOMA class
  \IfFileExists{parskip.sty}{%
    \usepackage{parskip}
  }{% else
    \setlength{\parindent}{0pt}
    \setlength{\parskip}{6pt plus 2pt minus 1pt}}
}{% if KOMA class
  \KOMAoptions{parskip=half}}
\makeatother
\usepackage{xcolor}
\usepackage{graphicx}
\makeatletter
\def\maxwidth{\ifdim\Gin@nat@width>\linewidth\linewidth\else\Gin@nat@width\fi}
\def\maxheight{\ifdim\Gin@nat@height>\textheight\textheight\else\Gin@nat@height\fi}
\makeatother
\setkeys{Gin}{width=\maxwidth,height=\maxheight,keepaspectratio}
\makeatletter
\def\fps@figure{htbp}
\makeatother
\newlength{\cslhangindent}
\newlength{\csllabelwidth}
\newlength{\cslentryspacingunit} % times entry-spacing
\newenvironment{CSLReferences}[2] % #1 hanging-ident, #2 entry spacing
 {% don't indent paragraphs
  \setlength{\parindent}{0pt}
  \ifodd #1
  \let\oldpar\par
  \def\par{\hangindent=\cslhangindent\oldpar}
  \fi
  \setlength{\parskip}{#2\cslentryspacingunit}
 }%
 {}
\usepackage{calc}

\ifLuaTeX
  \usepackage{selnolig}  % disable illegal ligatures
\fi
\IfFileExists{bookmark.sty}{\usepackage{bookmark}}{\usepackage{hyperref}}
\IfFileExists{xurl.sty}{\usepackage{xurl}}{} % add URL line breaks if available
\hypersetup{
  pdftitle={Establishing trust in automated reasoning},
  colorlinks=true,
  linkcolor={ForestGreen},
  filecolor={Maroon},
  citecolor={red},
  urlcolor={blue},
  pdfcreator={LaTeX via pandoc}}

\title{Establishing trust in automated reasoning}
\author{Konrad
Hinsen \\ {\small } \\ {\small } \\ {\small konrad.hinsen@cnrs.fr} \and  \\ {\small Centre
de Biophysique Moléculaire (UPR4301 CNRS)} \\ {\small Rue Charles
Sadron, 45071 Orléans Cédex 2,
France} \\ {\small } \and  \\ {\small Synchrotron SOLEIL, Division
Expériences} \\ {\small B.P. 48, 91192 Gif sur Yvette,
France} \\ {\small }} 
\date{}

\begin{document}
\maketitle
\begin{abstract}
Since its beginnings in the 1940s, automated reasoning by computers has
become a tool of ever growing importance in scientific research. So far,
the rules underlying automated reasoning have mainly been formulated by
humans, in the form of program source code. Rules derived from large
amounts of data, via machine learning techniques, are a complementary
approach currently under intense development. The question of why we
should trust these systems, and the results obtained with their help,
has been discussed by early practitioners of computational science, but
was later forgotten. The present work focuses on independent reviewing,
an important source of trust in science, and identifies the
characteristics of automated reasoning systems that affect their
reviewability. It also discusses possible steps towards increasing
reviewability and trustworthiness via a combination of technical and
social measures.
\end{abstract}

\hypertarget{introduction}{%
\section{Introduction}\label{introduction}}

Like all social processes, scientific research builds on trust. In order
to increase humanity's knowledge and understanding, scientists need to
trust their colleagues, their institutions, their tools, and the
scientific record. Moreover, science plays an increasingly important
role in industry and public policy. Decision makers in these spheres
must therefore be able to judge which of the scientific findings that
matter for them are actually trustworthy.

In addition to the trust-forming mechanisms present in social
relationships, the scientific method is built in particular on
\emph{transparency} and \emph{independent critical inspection}, which
serve to remove mistakes and biases in individual contributions as they
enter the scientific record. Ever since the beginnings of organized
science in the 17th century, with the appearance of learned societies
and the first scientific journals, researchers are expected to put all
facts supporting their conclusions on the table, and allow their peers
to inspect them for accuracy, pertinence, completeness, and bias. Since
the 1950s, following a sharp increase in the number of researchers and
the availability of photocopiers, critical inspection in the form of
\emph{peer review} has become an integral part of the publication
process {[}Spier 2002{]}. It is still widely regarded as a key criterion
for trustworthy results.

Over the last two decades, an unexpectedly large number of peer-reviewed
findings across many scientific disciplines have been found to be
irreproducible upon closer inspection {[}Baker 2016; Stodden et al.
2018; Colliard et al. 2022; Samuel and Mietchen 2024{]}. This so-called
``reproducibility crisis'' has shown that our practices for performing,
publishing, reviewing, and interpreting scientific studies are no longer
adequate in today's scientific research landscape, whose social,
technological, and economic contexts have changed dramatically. Updating
these processes is a major aspect of the nascent Open Science movement.

The topic of this article is a particularly important recent change in
research practices: the increasing use of automated reasoning. Computers
and software have led to the development of completely new techniques
for scientific investigation, and permitted existing ones to be applied
at larger scales and by a much larger number of researchers. In the
quantitative sciences, almost all of today's research critically relies
on computational techniques, even when they are not the primary tool for
investigation {[}Hettrick et al. 2014; Nangia and Katz 2017{]}.
Simulation, data analysis, and statistical inference have found their
place in almost every researcher's toolbox. Machine learning techniques,
currently under intense development, are likely to become equally
ubiquitous in the near future.

From the point of view of transparency and critical inspection, these
new tools are highly problematic. In philosophy of science, this lack of
transparence is referred to as \emph{epistemic opacity} {[}Humphreys
2009{]}. Among practitioners of computational science, it was a topic of
debate in the 1980s, when affordable desktop workstations first made
computation accessible to a larger number of scientists. It has been
summarized by Turkle {[}2009{]} as ``the tension between doing and
doubting'', i.e.~the tension between enthusiam for the new possibilities
and the doubts about the reliability of the very new and unproven
techniques. Turkle also describes the subsequent waning of doubt:
``Familiarity with the behavior of virtual objects can grow into
something akin to trusting them, a new kind of witnessing.'' Today, a
typical research paper reports the use of software uncritically, much
like the use of well-understood scientific instruments. There are rarely
any signs of reflection about the reliability of the software or about
its adequacy for the task being performed.

Automation bias {[}Parasuraman and Riley 1997{]}, i.e.~the propensity to
consider automated processes more reliable than human labor, is one
possible cause. Resignation to the impossibility of constructive doubt
is another one. The results produced by automated reasoning are often
neither obviously correct nor obviously wrong. In the absence of process
transparency, this leaves only uncritical acceptance or uncritical
rejection as possible reactions. Today's default is uncritical
acceptance.

The reproducibility crisis could have been a wakeup call, but the
contribution of automated reasoning to this crisis has not been widely
recognized. Several examples cited in this context involve faulty
software {[}Merali 2010{]}, e.g.~the retraction of papers in structural
biology due to a software bug {[}Miller 2006{]}, or the mistakes
discovered in a high-impact study on the relation between debt and
economic growth {[}Herndon et al. 2014{]}. These documented cases are
probably only the tip of the iceberg, given that detecting a mistake and
tracing it back to a software issue is an arduous task when all elements
have been published, and an impossible one otherwise. Pessimistic but
not unrealistic estimates suggest that most computational results in
science are to some degree wrong because of software defects {[}Soergel
2015; Thimbleby 2023{]}.

However, most high-profile cases of observed non-reproducibility involve
empirical research, e.g.~in medicine or psychology, in which automated
reasoning is not seen as a major aspect because it intervenes only in
the routine application of statistical software for data analysis.
Nevertheless, there are two aspects of this use of software that deserve
critical reflection. First, incorrect use of software is as much as
source of mistakes as defective software. Given the bad state of user
interface design and documentation for much scientific software,
incorrect use should be expected to happen. Second, inappropriate use is
an even more subtle source of errors. The mere existence of statistical
software has made statistical methods accessible to a much wider public.
Techniques that used to require collaboration with trained statisticians
are nowadays available as black-box tools to researchers who may not
even know what they would need to learn in order to use these techniques
correctly.

Given that it is impossible in practice to follow a computation step by
step, meaning processor instruction by processor instruction, in order
to verify its correctness, on what basis can scientists justify their
belief in its results? Durán and Formanek {[}Durán and Formanek 2018{]}
propose \emph{computational reliabilism} as a general framework. It
stipulates that a result can be trusted if the algorithm or piece of
software used to obtain it can be considered reliable. Reliability is
evaluated using various reliability indicators, which can be based on
characteristics of the source code but also on the observed behavior of
the algorithm and on the context of its actual use in scientifc practice
{[}Durán 2025{]}. The choice of suitable reliability indicators, as well
as the weight that should be attributed to each of them, remains an open
issue and should be expected to be context-dependent. From a pragmatic
rather than epistemologic point of view, reliabilism is very close to
how scientist actually reason about their software, once they start
reflecting on it.

From the perspective of computational reliabilism, critical inspection
has the goal of evaluating appropriate reliability indicators for each
piece of software. It can involve reading the source code and its
documentation, computing software quality metrics, testing programs on
well-known test cases, performing usability studies, etc. Given today's
complex software stacks, an exhaustive critical inspection is impossible
for the individual scientist wishing to deploy it, for lack of both time
and competence. The division of labor that makes the development of
complex software stacks possible requires a corresponding division of
labor in critical inspection. Individual components as well as larger
assemblies must be reviewed by independent experts, both in order to
provide feedback to software authors and in order to allow software
users to evaluate the reliability of their tools based on expert
judgment. An added complication is that reliability is contextual,
because the software's behavior can be appropriate for one use case but
unacceptable for another one. A review must therefore either be
exhaustive or clearly state the limits of its focus.

Open Science has made a first step towards dealing with automated
reasoning, in insisting on the necessity to publish scientific software,
and ideally making the full development process transparent by the
adoption of Open Source practices. While this level of transparency
greatly facilitates critical inspection, it is not sufficient. Someone
must actually look at this software with a critical eye, looking for
mistakes and for tacit assumptions that users need to know about. This
is not happening today, and we do not even have established processes
for performing such reviews {[}Hinsen 2025{]}. Moreover, as I will
explain later, much of today's scientific software is written in a way
that makes independent critical inspection particularly challenging if
not impossible. If we want scientific software to become trustworthy, we
therefore have to develop reviewing practices in parallel with software
architectures that make reviewing actually feasible in practice. And
where reviewing is not possible, we must acknowledge the fragile nature
of automated reasoning processes and make sure that everyone looking at
their results is aware of their uncertain reliability.

As for all research tools, it is not only the software itself that
requires critical inspection, but also the way the software is used in a
specific research project. Improper use of software, or
inappropriateness of the methods implemented by the software, can lead
to mistakes as well. However, it is much more difficult to detect with
today's minimal reporting practices. Moreover, the distinction between a
defect and inappropriate use is not as obvious as it may seem. A clear
distinction would require a well-defined interface between software and
users, much like a written contract. If the software's behavior deviates
from this contract, it's a defect. If the user's needs deviate from the
contract, it's inappropriate use. But such detailed contracts, called
\emph{specifications} in the context of software, rarely exist. The high
cost of writing, verifying, and maintaining specifications limits their
use to particularly critical applications. This means that reviewing the
use of scientific software requires particular attention to potential
mismatches between the software's behavior and its users' expectations,
in particular concerning edge cases and tacit assumptions made by the
software developers. They are necessarily expressed somewhere in the
software's source code, but users are often not aware of them.

The scientific requirement of \emph{independent} reviewing is related to
another aspect of automated reasoning that I will address, in particular
in my proposals for improving our current practices: the preservation of
epistemic diversity. As Leonelli has pointed out {[}Leonelli 2022{]},
the Open Science movement has so far largely neglected this aspect.
Epistemic diversity is about different perspectives and research
methodologies coexisting, enriching and critiquing each other.
Automation, be it in industry or in research, tends to reduce diversity
by favoring standardization as an enabler of economies of scale. In the
Open Science movement, this tendency is implicit in the quest for
reusability, one of the four FAIR principles {[}Wilkinson et al. 2016;
Barker et al. 2022{]}. Reusing someone else's code or data requires
adopting the authors' methodologies, and to some degree their general
perspective on the phenomenon under study. In the extreme case of a
single software package being used by everyone in a research community,
there is nobody left who could provide critical feedback.

This article has two main parts. In the first part (section
\ref{reviewability}), I look at the factors that make automated
reasoning more or less reviewable. It is a critical examination of the
state of the art in scientific software and its application, which
should help scientists to get a better grasp of how reliable automated
reasoning can be expected to be. In the second part (section
\ref{improving}), I consider how the reviewability of automated
reasoning can be improved, both through better reviewing processes and
by restructuring software for better reviewability.

\hypertarget{reviewability}{%
\section{Reviewability of automated reasoning
systems}\label{reviewability}}

Automated reasoning can play different roles in scientific research,
with different reliability requirements.\footnote{This is of course true
  for software in general, see e.g.~the discussion in {[}Shaw
  2022:22{]}.} The numerical preprocessing of observational data before
scientific analysis, nowadays often integrated into scientific
instruments, is an example where high reliability is required, because
its outputs are used without any further verification. On the other
hand, protein structure prediction by AlphaFold {[}Jumper et al. 2021{]}
is known to be unreliable, but it is nevertheless very useful if coupled
with experimental validation of its predictions {[}Nielsen 2023{]}.
Traditional computer simulation is often used similarly in biology as a
hypothesis generator whose outputs are subject to subsequent validation,
whereas in engineering, simulations of mechanical systems are routinely
performed to support critical decisions, thus requiring high
reliability.

What these examples illustrate is that tools, processes, and results in
science do not necessarily have to be perfectly reliable. Higher-level
validation processes act much like error correction protocols in
engineering. The coherence of multiple approaches to a question, coming
from different perspectives, is another higher-level source of
reliability, indicating robustness. This again illustrates the
importance of epistemic diversity that I have mentioned in the
introduction. What matters, however, is a clear understanding of the
reliability of individual scientific contributions, which in turn
requires a clear understanding of the reliability of the tools and
processes on which those contributions are based.

\begin{figure}
\hypertarget{fig:five-dimensions}{%
\centering
\includegraphics[width=0.7\textwidth,height=\textheight]{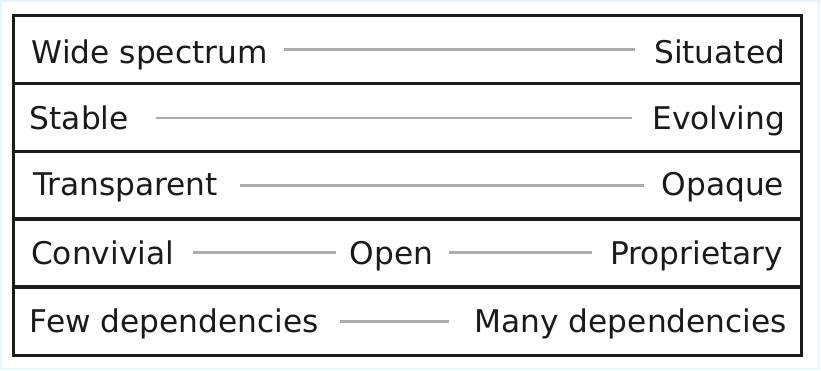}
\caption{The five dimensions of scientific software that influence its
reviewability.}\label{fig:five-dimensions}
}
\end{figure}

In this section, I discuss five characteristics (summarized in Fig.
\ref{fig:five-dimensions}) of automated reasoning systems that influence
how their reliability can be assessed by independent critical
inspection, which in the following I will call \emph{review} for
brevity. This use of \emph{review}, inspired by the tradition of
scientific peer review, should not be confused with the software
engineering technique of \emph{code review}, which is a quality control
step performed \emph{internally} by a development team. The term should
not be read either as a direct transposition of the pre-publication peer
review of papers by scientific journals, whose main output is an
accept-or-reject judgement. Reviewing can take various forms depending
on context. It can be performed by the researchers who wish to adopt
someone else's software tool, or by designated experts whose reports are
published.

Also for brevity, I will use the term \emph{software} instead of
``automated reasoning system'', extending its usual meaning to include
trained neural networks and other models obtained via machine learning
techniques. In science, software and machine learning differ in degree
rather than in kind, as already long before machine learning, many
computational models contained a large number of parameters fitted to
large datasets.

Recently, Hasselbring \emph{et al.} {[}2025{]} have published a
multidimensional categorization of research software that doesn't
address reviewability, but has categories that overlap partially with
the ones I define in the following. Readers unfamiliar with the
diversity of research software would profit from reading their analysis
before continuing here.

\begin{figure}
\hypertarget{fig:software-stack}{%
\centering
\includegraphics[width=0.7\textwidth,height=\textheight]{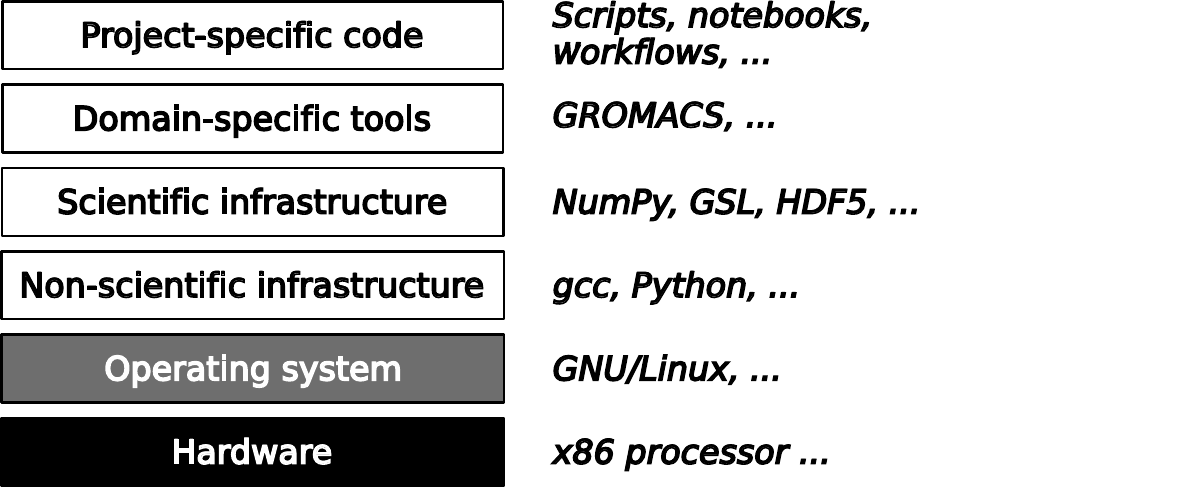}
\caption{A typical software stack as used in a research
project}\label{fig:software-stack}
}
\end{figure}

Fig. \ref{fig:software-stack} shows the layers of a typical software
stack for a research project, which I will refer to in the following.
The top layer is software written specifically for a research project.
This layer usually contains small items: scripts, notebooks, or
workflows. They use software components from the layer below, which I
have labelled ``domain-specific tools''. This domain-specific layer
contains software that implements the computational models and methods
used by a research community. Many ongoing discussions of scientific
software, in particular concerning its sustainability {[}Hettrick
2016{]}, concentrate on this layer, but are not always explicit about
this focus.

The two levels further down contain infrastructure software, meaning
software that the domain-specific layer builds on but which
computational scientists are often only vaguely aware of. They may for
example know that their software is written in the C language, without
knowing which precise version of which C compiler was used to compile
it. I have divided infrastructure into two layers: a general one that
computational science shares with other application domains of
computing, and a scientific layer for software that is written
explicitly to support science and engineering.

\hypertarget{wide_spectrum_vs_situated}{%
\subsection{Wide-spectrum vs.~situated
software}\label{wide_spectrum_vs_situated}}

Wide-spectrum software provides fundamental computing functionality to a
large number of users. In order to serve a large user base, it addresses
a wide range of application scenarios, each of which requiring only a
part of the software's functionality. Word processors are a well-known
example: a package like \href{https://www.libreoffice.org/}{LibreOffice}
can be used to write a simple letter, but also a complex book.
LibreOffice has huge menus filled with various functions, of which most
users only know the handful that matters to them. General-purpose large
language models are another example of wide-spectrum software.

Situated software (a term introduced by Shirky {[}2004{]}) is software
written for a specific use case or a narrow user group. It addresses a
specific need very well, but is not transferable to other application
scenarios. Spreadsheets are usually situated, as are games, and many
shell scripts.

A useful numerical proxy for estimating a software package's location on
this scale is the ratio of the number of users to the number of
developers, although there are exceptions. Games, for example, are
situated software with few developers but many users.

In scientific computing, the wide-spectrum end of the scale is well
illustrated by mathematical libraries such as
\href{https://www.openblas.net/}{BLAS} or visualization libraries such
as \href{https://matplotlib.org/}{matplotlib}, which provide a large
collection of functions from which application developers pick what they
need. They are found mostly in the infrastructure layer of Fig.
\ref{fig:software-stack}, but also in the domain-specific layer. At the
situated end, we have mainly the project-specific code at the top of
Fig. \ref{fig:software-stack}: code snippets and scripts that generate
the plots shown in a paper, or computational notebooks and workflows.
Most of the domain-specific layer lies in between these extremes; see
section \ref{gromacs} below for a case study.

Reviewing wide-spectrum software represents a major effort, because of
its size and functional diversity. Moreover, since wide-spectrum
software projects tend to be long-lived, with the software adapting to
new use cases and new computing platforms, its critical examination must
be an ongoing process as well. On the other hand, this effort can be
considered a good investment, because of the large user base such
software has.

Situated software is usually smaller and simpler, which makes it easier
to understand and thus to review. However, its evaluation can only be
done in the specific context for which the software was written. This
suggests integrating it into the existing scientific peer reviewing
process, along with papers and other artifacts that result from a
research project.

It is the intermediate forms of software that are most difficult to
review. Domain tools and libraries are too large and complex to be
evaluated in a single session by a single person, as is expected in
today's journal peer review process. However, they don't have a large
enough user base to justify costly external audits, except potentially
in contexts such as high-performance computing where the importance of
the application and the high cost of the invested resources also justify
more careful verification processes.

\hypertarget{stability}{%
\subsection{Stable vs.~evolving software}\label{stability}}

Stable software is developed and maintained with the goal of providing a
reliable tool. Signs of stability in software are its age, a clear
definition of its purpose, ideally in the form of a specification,
respect of standards, respect of software engineering practices,
detailed documentation, and the absence of compatibility-breaking
changes. The Linux kernel and the text editor Emacs are examples of very
stable software.

Stable does not mean static. It is common for stable software to grow by
addition of new functionality. It is also common for stable software to
be ported to new computational environments. Stability is evaluated from
the user's point of view, not by counting changes to the source code.
Both the documented behavior of the software and the tacit knowledge
that users acquire over the years should be stable.

Evolving software is either software in early stages of development, for
which design and implementation decisions are still revised from time to
time, or software whose goals change over time. One important category
of evolving software is prototype software, i.e.~software created to
test new ideas, be they technical (software architecture etc.) or
related to the application domain. For software projects that adopt
\href{https://semver.org/}{semantic versioning}, the major version
number 0 indicates prototype status.

Another important category is software that is continuously adapted to
changing requirements. This has become common over the last two decades,
in particular within Open Source communities. As an illustration of this
rising attitude, consider programming languages: all languages with a
written standard and multiple implementations are more than 20 years old
(but the standards are still updated from time to time), whereas
programming languages published more recently have a single
authoritative implementation that acts as an evolving \emph{de facto}
specification of their behavior. This does not prevent them from being
stable, by consensus of the developer community, as e.g.~for the Clojure
programming language {[}Pote 2025{]}, but stability cannot be taken for
granted, even for a 30 year old language such as Python. Unfortunately,
developers usually make no clearly visible statement about aiming for
stability or for continuous adaptation.

Research software can be anything from very stable to rapidly evolving.
Many domain-specific software packages (see Fig.
\ref{fig:software-stack}) contain both stable and evolving parts. For
their developer communities, which are typically small, keeping stable
and evolving parts separate is neither easy nor particularly important,
because the potential users for both parts are the same. Infrastructure
ought to be stable in order to ensure a solid foundation for a software
stack. This is generally true for infrastructure software having its
roots in the 20th century. Many more recent software packages are
\emph{de facto} used as infrastructure in spite of being evolving. See
section \ref{numpy} below for a case study.

For reviewing, evolving software is mainly an economic issue. Reviewing
takes time, and the reviews must evolve at the same pace as the software
itself. But from the wider perspective of reliability and trust,
evolving software raises the more fundamental issue of what reliability
even means for a piece of software whose intended behavior is in flux.
Building trust of any kind takes time. It requires repeated probing with
dominantly positive returns. People will never see software as reliable
if it evolves faster than they can build trust in it. This implies that
reliability is relational, i.e.~not an absolute characteristic of a
piece of software but a judgement made from a specific perspective. The
same software package can be seen as reliable by a power user, but as
unstable by an occasional user.

Finally, evolving software requires more attention from its users, who
must ensure that their knowledge about the software is up to date. The
helpful hint found on your colleague's Web site may no longer be
appropriate. Your own well-documented use case from a few years ago may
no longer be a good template to start a new project. Evolving software
thus comes with a higher cognitive load and an increased risk of
inappropriate use.

\hypertarget{opacity}{%
\subsection{Transparent vs.~opaque software}\label{opacity}}

Transparent software is software whose behavior is comprehensible and
verifiable by its users. Transparency differs from openness in that it
is about the behavior of a program, whereas openness is about its source
code. In a word processor, or a graphics editor, many user actions, such
as pasting in a piece of text or an image, produce an immediately
visible result that the user can easily predict and verify. In contrast,
opaque software performs complex operations and produces output whose
correctness cannot be assessed with reasonable effort. Large language
models are an extreme example: their output is impossible to predict and
verify. Note that this notion of opacity is not identical but similar to
the epistemic opacity I discussed in the introduction.

Strictly speaking, transparency is not a characteristic of a piece of
software, but of a computational task. A single piece of software may
contain both transparent and opaque functionality. In the word processor
example, inserting a character is highly transparent, whereas changing
the page layout is more opaque, creating the possibility of subtle bugs
whose impact is not readily observable. I use the term ``opaque
software'' as a shorthand for software implementing at least one major
opaque operation.

In scientific software, transparency often comes from a simple
specification for which multiple implementations are available. Examples
are solvers for partial differential equations in fluid dynamics,
quantum chemistry computations, or common numerical libraries for linear
algebra or fast Fourier transforms. The existence of multiple
implementations of precisely the same task makes it possible to verify
results from one implementation by comparing to the output of other
implementations. Moreover, the simplicity of the specification makes it
possible to perform tests on inputs for which the exact solution is
known by different means.

Most scientific software is closer to the opaque end of the spectrum.
Even highly interactive software, for example in data analysis, performs
non-obvious computations, yielding output that an experienced user can
perhaps judge for plausibility, but not for correctness. Statistical
inference processes are a good example. As a rough guideline, the more
scientific models or observational data have been integrated into a
piece of software, becoming a tacit part of its specification, the more
opaque the behavior of the software is to its users.

It is much easier to evaluate the reliability of transparent software,
because its expected output can be obtained independently. Reviewing
transparent software is therefore easier, to the point that it can often
be done by the user rather than by an expert. Moreover, when users can
understand and judge the results produced by a piece of software, even a
very weak reliability indicator such as popularity becomes
useful.\footnote{A famous dictum in software engineering, often referred
  to as \href{https://en.wikipedia.org/wiki/Linus\%27s_law}{``Linus'
  law''}, states that ``given enough eyeballs, all bugs are shallow''.
  However, this can only work if the many eyeballs are sufficiently
  trained to spot problems, meaning that users of opaque software don't
  qualify.}

The more opaque a computation is, the more important its documentation
becomes. Reliability can only be judged for the performance of a
well-defined task. If users cannot know what the software promises to do
exactly, nor what its limitations are, then they cannot perform any task
reliably using this software. This is currently a much discussed issue
with machine learning models {[}e.g. Gunning et al. 2019; Longo et al.
2024{]}, but it is not sufficiently recognized that traditional computer
software can be just as opaque from a user's point of view, if source
code is the only available documentation of its behavior.

\hypertarget{convivial-vs.-proprietary-software}{%
\subsection{Convivial vs.~proprietary
software}\label{convivial-vs.-proprietary-software}}

Convivial software {[}Kell 2020{]} is named in reference to Ivan
Illich's book ``Tools for conviviality'' {[}Illich 1973{]} It draws on
the root meaning of the latin \emph{convivium}, which is ``living
together''. Convivial technology as described by Illich is shaped by the
community of its users according to their evolving needs, rather than
being controlled by specialists outside of the community. Convivial
software aims at augmenting its users' agency over their computations by
actively encouraging them to adapt the software to their needs.
\href{https://malleable.systems/}{Malleable software} is a similar
concept, as is re-editable software, a term introduced
\href{https://web.archive.org/web/20101203111941/https://www.informit.com/articles/article.aspx?p=1193856}{by
Donald Knuth in an interview} in opposition to reusable,
i.e.~off-the-shelf, software {[}Hinsen 2018a{]}.

In contrast, proprietary software offers users fixed functionality and
therefore limited agency. At first sight this looks like users should
always prefer convivial software, but agency comes at a price: users
have to invest more effort and assume responsibility for the final
software assembly that they have made themselves. In fact, conviviality
is all about diminishing the gap between software developers and
software users. Just like most people prefer to choose from a range of
industrially-made refrigerators, rather than modify a generic design
precisely to their needs, most computer users are happy to use
ready-made e-mail software rather then assembling their own from items
in a toolkit.

It is important to understand that conviviality is not a characteristic
of software as a digital artifact (e.g.~the contents of a repository on
GitHub), but of a software ecosystem as a socio-technical system. A
well-known convivial ecosystem is the collection of Unix command line
tools, which users can freely combine in shell scripts and complement by
adding their own tools. It doesn't make much sense to call individual
command line tools such as \texttt{ls} or \texttt{grep} convivial, but
it does make sense to say that they are designed to be part of a
convivial ecosystem. A necessary but not sufficient condition for
fitting into a convivial ecosystem is transparency (see section
\ref{opacity} and sections \ref{numpy} and \ref{gromacs} for relevant
case studies).

In the academic literature on software enginering, convivial software is
discussed with the focus on its users becoming developers, most commonly
referred to as \emph{end user programmers} {[}Nardi 1993; Ko et al.
2011{]}. Shaw recently proposed the less pejorative term
\emph{vernacular developers}, which she defines as ``people who are not
professionally trained as programmers {[}and who{]} are creating and
tailoring software as a means to achieve goals of their own'' {[}Shaw
2022{]}. The subfield of end user software engineering aims at providing
vernacular developers with methods and tools to improve the quality of
their software, recognizing that the methods and tools designed for
software professionals are usually not adapted to their needs.

The risk of non-convivial technology, which Illich {[}1973{]} discusses
in detail, is that widespread adoption makes society as a whole
dependent on a small number of people and organizations who control the
technology. This is exactly what has happened with computing technology
for non-professional use, such as personal computers and smartphones.
You may not want to let corporations spy on you via your smartphone, but
the wide adoption of these devices means that you are excluded from more
and more areas of social life if you decide not to use one. Many
research communities have fallen into this trap as well, by adopting
proprietary infrastructure, such as
\href{https://www.mathworks.com/products/matlab.html}{MATLAB}, or Open
Source infrastructure over which they have no control, such as Python,
as a foundation for their computational tools and models. A convivial
infrastructure would have preserved these researchers' autonomy to shape
the software according to the needs of their research, at the cost of
much stronger participation in its development and maintenance.

In between convivial and proprietary software, we have Free, Libre, and
Open Source software (FLOSS). Historically, the Free Software movement
was born in the 1980s in academia {[}Gonzalez-Barahona 2021{]}. Much
software was distributed with its source code, even in commercial
settings. Most computer users in this environment needed to do some
programming of their own in order to get anything done. In other words,
software was much more convivial than what we have today. It is the
arrival of proprietary software in their lives, exemplified by the
frequently cited proprietary printer driver at MIT {[}2002{]}, that
pushed them towards formalizing the concept of Free Software in terms of
copyright and licensing.

With the enormous complexification of software over the following
decades, a license is no longer sufficient to keep software convivial in
practice. The right to adapt software to your needs is of limited value
if the effort to do so is prohibitive. Software complexity has led to a
creeping loss of user agency, to the point that even building and
installing Open Source software from its source code is often no longer
accessible to non-experts because of the complexity of the build tools.
An experience report on building the popular machine learning library
\href{https://pytorch.org/}{PyTorch} from source code nicely illustrates
this point {[}Courtès 2021{]}. The authors of PyTorch explain how you
can install the precompiled versions they provide (via Python's
\texttt{pip} utility), but they do not explain how to compile the code
on your own, as you would have to do to fix bugs or tweak some
functionality. This makes it practically impossible for most PyTorch
users to adapt the code to their needs. The work by Courtès {[}2021{]}
amounts to reverse-engineering the build process and publishing it as a
build recipe for \href{https://guix.gnu.org/}{Guix}, a package manager
designed for reproducible builds from source code.

Conviviality has become a marginal subject in the FLOSS movement, with
the Free Software subcommunity pretending that it remains ensured by
copyleft licenses and much of the Open Source subcommunity not caring
about it. It survives mainly in communities whose technology has its
roots in the 1970s and 1980s, when ideas such as Alan Kay's Dynabook
{[}Kay 1972{]}, a device for which even children could create software
for their own use, were active research topics. Examples are programming
systems inheriting from Smalltalk
(e.g.~\href{http://squeak.org/}{Squeak},
\href{http://pharo.org/}{Pharo}, and \href{https://cuis.st/}{Cuis}), or
the programmable text editor
\href{https://www.gnu.org/software/emacs/}{GNU Emacs}. These systems
encourage modification via built-in tools for discovering functionality
and for locating the parts of the code that implement it.

In scientific computing, there is a lot of diversity on this scale.
Fully proprietary software is common, but also variants that do allow
users to look at the source code, but don't allow them to compile it, or
don't allow the publication of reviews. In computational chemistry, the
widely used Gaussian software is an example for such legal constraints
{[}Hocquet and Wieber 2017{]}. FLOSS has been rapidly gaining
popularity, and receives strong support from the Open Science movement
(see e.g.~the Second French National Plan for Open Science {[}French
Ministry for Higher Education, Research, and Innovation 2021{]} that has
a whole chapter dedicated to software). Somewhat surprisingly, the move
beyond FLOSS to convivial software is hardly ever envisaged, in spite of
it being aligned with the pre-digital practices of scientific research:
the main intellectual artifacts of science, i.e.~theories and models,
have always been convivial.

Concerning reviewing, the convivial-to-open part of the scale is similar
to the situated-to-wide-spectrum scale: convivial software is easier to
understand and therefore easier to review, but each specific adaptation
of convivial software requires its own review, whereas open but not
convivial software makes reviewing a better investment of effort.
Proprietary software is harder to review, because only its observed
behavior and its documentation are available for critical inspection,
but not its source code.

\hypertarget{resources}{%
\subsection{Size of the minimal supporting
environment}\label{resources}}

Each piece of software requires a computational environment, consisting
of a computer and of other pieces of software. In terms of the software
stack illustrated by Fig. \ref{fig:software-stack}, the computational
environment for a piece of software consists is situated in the layers
below its own. The importance of computational environments is not
sufficiently appreciated by most researchers today, which is the main
cause of widespread computational irreproducibility {[}Stodden and
Miguez 2013; Hinsen 2020; Wang et al. 2021{]}.

It is the computational environment that defines what a piece of source
code actually does. The meaning of a Python script is defined by the
Python interpreter. The Python interpreter is itself a piece of software
written in the C language, and therefore the meaning of its source code
is defined by the C compiler, and by the processor which ultimately
executes the binary code produced by the C compiler. As an illustration
for the importance of the computational environment, it is an easy
exercise to write a Python script that produces different results when
run with version 2 or version 3 of the Python interpreter, exploiting
the different semantics of integer division between the two versions.

In addition to this semantic importance of computational environments,
reviewability implies a pragmatic one: reviewers of software or its
results need access to an adequate hardware and software environment in
order to perform their review. Scientific computing mostly relies on
commodity hardware today, with two important exceptions: supercomputers
and Graphical Processing Units (GPUs). Supercomputers are rare and
expensive, and thus not easily accessible to a reviewer. GPUs are
evolving rapidly, making it challenging to get access to an identical
configuration for reviewing. Supercomputers often include GPUs,
combining both problems. Resource access issues are manageable for
wide-spectrum software if they are deemed sufficiently important to
warrant the cost of performing audits on non-standard hardware.

Software environments have been recognized as highly relevant for
automated reasoning in science and beyond, and are the subject of active
research {[}e.g. Malka et al. 2024; Bilke et al. 2025{]}. They play a
key role in computational reproducibility, but also for privacy and
security, which are the prime motivations for the
\href{https://reproducible-builds.org/}{Reproducible Builds} movement
\footnote{Ken Thompson's Turing Award Lecture ``Reflections on Trusting
  Trust'' {[}Thompson 1984{]} is an early and very readable discussion
  of the security implications of computational environments.}. The
issues of managing software environments are now well understood, and
two software management systems (\href{https://nixos.org/}{Nix} and
\href{https://guix.gnu.org/}{Guix}) implement a comprehensive solution.
However, they are so far used only by a small minority of researchers.
In addition to ease of use issues, which could be overcome with time and
investements, a major obstacle is that such management systems must
control the complete software stack, which excludes the use of popular
proprietary platforms such as Windows \footnote{Windows is a trademark
  of the Microsoft group of companies.} or macOS \footnote{macOS is a
  trademark of Apple Inc., registered in the U.S. and other countries
  and regions.}.

Assuming that the proper management of scientific software envronments
will be achieved not only in theory, but also in practice, it is the
size of this environment that remains a major characteristic for
reviewability. The components of the computational environment required
by a piece of software are called its \emph{dependencies} in software
engineering. This term expresses their importance very well: every
single quality expected from a software system is influenced by the
corresponding quality of the components that enter in its construction.
For example, no software can be more stable than its dependencies,
because of the risk of software collapse {[}Hinsen 2019{]}. Reviewing
software therefore requires an examination of its dependencies as well.
This can become an obstacle for software that has hundreds or even
thousands of dependencies. Imprecise dependency references
(e.g.~``version 2 or later of package X'') are a problem as well,
because they require reviewers to emit a judgement on a potentially
large number of versions of a package, and to extend their trust in them
to an indefinite future.

\hypertarget{analogies-in-experimental-and-theoretical-science}{%
\subsection{Analogies in experimental and theoretical
science}\label{analogies-in-experimental-and-theoretical-science}}

For developing a better understanding of the reviewability
characteristics described above, it is helpful to consider analogies
from the better understood experimental and theoretical techniques in
scientific research. In particular, it is helpful to examine where such
analogies fail due to the particularities of software.

Experimental setups are situated. They are designed and constructed for
a specific experiment, described in a paper's methods section, and
reviewed as part of the paper review. Most of the components used in an
experimental setup are mature industrial products, ranging from
commodities (electricity cables, test tubes, etc.) to complex and
specialized instruments, such as microscopes and NMR spectrometers.
Non-industrial components are occasionally made for special needs, but
this is discouraged by their high manufacturing cost. The use of
prototype components is exceptional, and usually has the explicit
purpose of testing the prototype. Some components are very transparent
(e.g.~electricity cables), others are very opaque (e.g.~NMR
spectrometers). The equivalent of the computational environment is the
physical environment of the experimental setup. Its impact on the
observations tends to be well understood in the physical sciences, but
less so in the life sciences, where it is a common source of
reproducibility issues (e.g. {[}Kortzfleisch et al. 2022{]} or
{[}Georgiou et al. 2022{]}).

The main difference to software is thus the much lower prevalence of
prototype components. A more subtle difference between instruments and
software is that the former are carefully designed to be robust under
perturbations, whereas computation is chaotic {[}Hinsen 2016{]}. A
microscope with a small defect may show a distorted image, which an
experienced microscopist will recognize when evaluating the microscope.
Software with a small defect, on the other hand, can introduce
unpredictable errors in both kind and magnitude, which neither a domain
expert nor a professional programmer or computer scientist can diagnose
easily if the software is opaque. This is a consequence of software
being crafted out of a substrate (Turing-complete languages) that is
much less constraining than the physical substrates from which
instruments are made. The increasing integration of computers and
software into scientific instruments may lead to experimental setups
becoming less robust as well in the future.

Analogies with pre-digital scientific models and theories are
instructive as well. Wide-spectrum theories exist in the form of
abstract reasoning frameworks, in particular mathematics. The analogue
of situated software are concrete models for specific observational
contexts. In between, we have general theoretical frameworks, such as
evolutionary theory or quantum mechanics, and models that intentionally
capture only the salient features of a system under study, pursuing
understanding rather than precise prediction. Examples for the latter
are the Ising model in physics or the Lotka-Volterra equations in
ecology.

Abstract frameworks and general theories are the product of a long
knowledge consolidation process, in which individual contributions have
been reviewed, verified on countless applications, reformulated from
several perspectives, and integrated into a coherent whole. This process
ensures stability, transparency, and conviviality in a way that has so
far no equivalent for software.

Opacity is an issue for theories and models as well: they can be so
complex and specialized that only a handful of experts understand them.
It also happens that people apply such theories and models
inappropriately, for lack of sufficient understanding. However,
automation via computers has amplified the possibility to deploy opaque
sets of rules so much that it makes a qualitative difference: scientists
can nowadays use software whose precise function they could not
understand even if they dedicated the rest of their career to it.

The computational environment for theories and models is the people who
work with them. Their habits, tacit assumptions, and metaphysical
beliefs play a similar role to hardware and software dependencies in
computation, and they are indeed also a common cause of mistakes and
misunderstandings.

\hypertarget{casestudies}{%
\subsection{Case studies}\label{casestudies}}

In order to illustrate the five dimensions described above, I will
present a few case studies on software that I know well, from having
used it in my own research work.

\hypertarget{numpy}{%
\subsubsection{NumPy}\label{numpy}}

NumPy {[}Harris et al. 2020{]} is the foundational library of the
Scientific Python ecosystem. It adds an efficient array data structure
to the Python language, which serves both for efficient manipulation of
scientific data from Python and for interfacing Python with lower-level
languages such as C or Fortran. NumPy is an interesting case study
because of its status as widely used infrastructure. It is also an
example I know exceptionally well because I was involved in its early
development stages {[}Dubois et al. 1996{]}.

\textbf{Scope:} NumPy is used in all disciplines of quantitative
science, for data of very diverse types, including geometric objects,
pixel images, and neural networks. It is therefore clearly a
wide-spectrum software package.

\textbf{Openness:} NumPy is Open Source software. It is the foundation
of an ecosystem that is not convivial, being much too complex for its
users to understand or modify. However, its transparency (see below)
would make it a candidate for a foundation of a convivial research
software ecosystem, if it could be made more stable (see below).

\textbf{Stability:} The stability status of NumPy is what makes it an
interesting case study. NumPy grew out of its predecessor
\emph{Numerical Python}, whose development started in the 1990s
{[}Dubois et al. 1996{]}. It was very stable in its first decade, with
the library and its dependees growing but maintaining backward
compatibility. In the 2000s, the popularity of the scientific Python
ecosystem grew rapidly. In 2006, NumPy was published as a major redesign
that unified the original Numerical Python with its similar but not
fully compatible offspring \emph{numarray}. At this occasion, numerous
changes were made to the API to ease the transition from Matlab to
Python, corresponding to the needs of a significant part of the freshly
recruited community members. The old API was maintained under a
different name in order to make migration from Numerical Python to NumPy
straightforward and risk-free. The intention was to let the two API
versions live side by side indefinitely. But another few years later,
after more growth, the old API was declared obsolete and then removed,
by younger maintainers who had no memories of it anyway. This broke some
of the oldest scientific software applications that had been written in
Python. However, their developers and users were no longer
representative for the community, whose culture had shifted to valuing
innovation over stability, in alignment with the wider technology
culture {[}Hinsen 2024b{]}. Since then, NumPy has regularly introduced
breaking changes in order to make way for what the developers of the
moment considered improvements, culminating in a recent major-version
update to NumPy 2 that comes with an extensive
\href{https://numpy.org/doc/stable/numpy_2_0_migration_guide.html}{migration
guide}.

\textbf{Transparency}: Basic array operations are very transparent.
However, there are many of them, and some have non-trivial edge cases.
Verifying the correct behavior of all operations is not difficult but
laborious. An extensive collection of unit tests with good coverage
takes care of that.

\textbf{Dependencies:} Identifying the dependencies of NumPy is not an
easy exercice. The installation instructions claim that ``the only
prerequisite for installing NumPy is Python itself.'' Howver, the
detailed instructions for compiling the source code add compilers for C
and C++, the libraries BLAS and LAPACK, and pkg-config. All of these
packages have their own dependencies as well. The complete dependency
graph for NumPy 2.3.1 as recorded by the
\href{https://guix.gnu.org/}{package manager Guix} in commit
\texttt{da04b3bb42d76e7af9d5eb344cfde2350e9bb3c1} contains 239 items
(including NumPy itself) \footnote{To reproduce this number, run
  \texttt{guix\ time-machine\ -\/-commit=da04b3bb42d76e7af9d5eb344cfde2350e9bb3c1\ -\/-\ graph\ python-numpy\ \textbar{}\ grep\ label\ \textbar{}\ wc\ -l}}.
Most of these packages are well-known and widely used stable elements of
common computational infrastructure, the notable exception being the
Python language, which has been evolving in a way similar to NumPy
itself.

Its evolving nature makes NumPy difficult to review, but on the other
hand its transparency makes reviewing less important, because software
engineering practices already provide sufficient reliability indicators.
However, the evolving nature of NumPy (and its main dependency, Python)
is a significant obstacle to reviewing software that builds on it. That
is an enormous amount of software across many disciplines. NumPy is
\emph{de facto} part of today's scientific computing infrastructure.
Infrastructure ought to be stable, but NumPy and Python are not.

A major contribution to this unsatisfactory situation is the lack of
support for computational infrastructure in general. Infrastructure
maintenance requires institutions that oversee and fund it from a
long-term perspective, and with a governance model in which all types of
users are represented. Such institutions do not exist today. As a
consequence, infrastructure such as NumPy is developed according to
shifting criteria that represent the interests of whoever happens to be
funding the current generation of developers.

\hypertarget{gsl}{%
\subsubsection{GNU Scientific Library (GSL)}\label{gsl}}

The GNU Scientific Library {[}Galassi 2009; 2025{]} implements widely
used algorithms for numerical mathematics in the C language, with
interfaces for many other languages. It has been under continuous
development since 1996. GSL is an interesting case study in comparison
with NumPy, because it is also infrastructure software but very stable
and more convivial.

\textbf{Scope:} GSL is a wide-spectrum library, containing functionality
of interest to many domains of quantitative science.

\textbf{Openness:} GSL is Free Software (GPL license), and thus open. It
is even halfway towards convivial. Being a collection of independent or
loosely coupled modules, it makes modifying a single module quite
accessible. When used as a component of a software stack, GSL is
typically situated in a lower layer, making it just as difficult to
replace with a modified version as any other lower-layer component. But
it is feasible to extract a GSL module and add a modified copy of it,
under a different name, to the software stack again.

\textbf{Stability:} A look at the
\href{https://cgit.git.savannah.gnu.org/cgit/gsl.git/tree/NEWS}{summary
of changes} shows that GSL is very stable. Changes fix bugs, add new
functionality, and add support for more platforms.

\textbf{Transparency:} Being a collection of implementations of
well-known algorithms of moderate complexity, GSL is quite transparent.

\textbf{Dependencies:} The only dependency for using GSL is a C
compiler. The
\href{https://en.wikipedia.org/wiki/Basic_Linear_Algebra_Subprograms}{BLAS}
implementation included with GSL can be replaced by a more efficient
one, but this is optional. Modifying and then rebuilding GSL, including
its documentation, requires five few well-known and stable build tools:
\href{https://www.gnu.org/software/autoconf/}{autoconf},
\href{https://www.gnu.org/software/automake/}{automake},
\href{https://www.gnu.org/software/libtool/}{libtool},
\href{https://www.gnu.org/software/texinfo/}{texinfo}, and
\href{https://www.sphinx-doc.org/}{Sphinx}.

Stable, transparent, and open software with few dependencies are easily
reviewable. See section \ref{reviewing} for possible reviewing
processes.

\hypertarget{gromacs}{%
\subsubsection{GROMACS}\label{gromacs}}

GROMACS {[}Abraham et al. 2015; Abraham et al. 2025{]} is a collection
of programs for performing simulations of molecular systems, with a
focus on biological macromolecules and on very large simulations
requiring high-performancs computing (HPC) resources. It is a typical
example of domain-specific scientific software: well-known among
practitioners of biomolecular simulations, but almost unknown outside of
this community.

\textbf{Scope:} GROMACS is neither wide-spectrum nor situated, but
somewhere in between.

\textbf{Openness:} GROMACS is Open Source software. It is a small
ecosystem by itself since it consists of multiple programs designed to
work together. This ecosystem is, however, not convivial. The source
code is far too complex for the typical GROMACS user to understand, and
some of the individual tools are opaque. Non-conviviality is confirmed
by the observation that the developer community is a small subset of the
user community.

\textbf{Stability:} GROMACS development follows a well-documented
release cycle, with one major release per year and several bug-fix
releases over a period of two years. This means that at any given time,
two major releases are supported by bug fixes. Major releases mostly add
functionality, but also improve performance and add support for new
computing systems, which is an important aspect for HPC where
performance requires adapting to specific processors, memory layouts,
and communication networks. The basic functionality is intended to be
stable, but its implementation is not because it evolves due to ongoing
optimization and porting efforts. Whether or not the basic functionality
actually \emph{is} stable is hard to decide because of opacity (see
below). GROMACS is therefore evolving software.

\textbf{Transparency:} The GROMACS code is a highly optimized
implementation of several computational models for molecular
interactions (called \emph{force fields}) and several simulation
methods. These ingredients are inextricably interwoven in the code
because of the primacy of performance. The implemented models and
methods are described and discussed in the scientific literature, and
also implemented in other software. However, no precise and complete
description exists anywhere outside of software source code. There is no
point in even asking if GROMACS implements any such model or method
correctly, as there is no precise notion of correctness. For this
reason, GROMACS is highly opaque.

\textbf{Dependencies:} The core dependencies are a C/C++ compiler, the
\href{https://fftw.org/}{FFTW} library for computing Fourier transforms,
an implementation of the
\href{https://en.wikipedia.org/wiki/Basic_Linear_Algebra_Subprograms}{BLAS}
and \href{https://en.wikipedia.org/wiki/LAPACK}{LAPACK} specifications
for linear algebra operations, plus the build system
\href{https://cmake.org/}{cmake}. All of these packages, as well as
their respective dependencies, are well-known stable elements of
computational infrastructure. However, GROMACS depends critically on
properties of specific compilers on specific machines, due to its
high-performance focus. It also has additional dependencies for specific
types of computers, which include proprietary libraries for using nVIDIA
GPUs. A quick glance at the
\href{https://manual.gromacs.org/2025.3/install-guide/index.html}{GROMACS
installation instructions} gives an idea of the potential complexity of
its dependency graph.

GROMACS is a good example for scientific software that is impossible to
review in its entirety, due to the combination of being opaque,
evolving, and having complex and platform-dependent dependencies. At
best, a review could cover a specfic version running on a specific
platform used for a narrowly defined type of application. Opacity and
instability also increase the risk of inappropriate use by scientists
who have an insufficient or outdated mental model of the software's
operation. I have personally witnessed several cases in which
inappropriate use of GROMACS led to obviously wrong results, which
suggests that there are also cases that go unnoticed.

Given today's development tools and best practices in software
engineering, plus the focus on performance, it is impossible to make the
software more reviewable. For potential measures to design more
reviewable molecular simulation software, see sections \ref{situated},
\ref{explainability} and \ref{specifications}.

\hypertarget{improving}{%
\section{Improving the reviewability of automated reasoning
systems}\label{improving}}

The analysis presented in the previous section can by itself improve the
basis for trust in automated reasoning, by providing a vocabulary for
discussing reviewability issues. Ensuring that both developers and users
of scientific software are aware of where the software is located on the
different scales I have described makes much of today's tacit knowledge
about scientific software explicit, avoiding misplaced expectations.

\begin{figure}
\hypertarget{fig:four-measures}{%
\centering
\includegraphics[width=0.7\textwidth,height=\textheight]{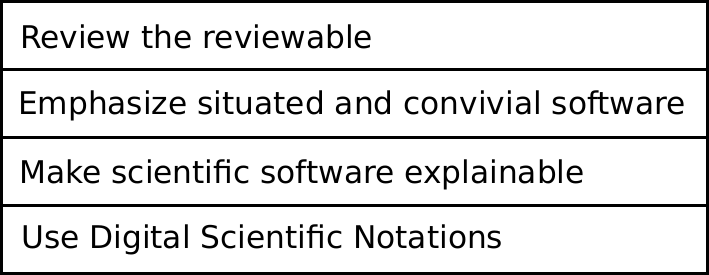}
\caption{Four measures that can be taken to make scientific software
more trustworthy.}\label{fig:four-measures}
}
\end{figure}

However, scientists can also work towards improving their computational
practices in view of more reviewable and thus ultimately more reliable
results. These improvements include both new reviewing processes,
supported by institutions that remain to be created, and new software
engineering practices that take into account the specific roles of
software in science, which differ in some important respects from the
needs of the software industry. The four measures I will explain in the
following are summarized in Fig. \ref{fig:four-measures}.

\hypertarget{reviewing}{%
\subsection{Review the reviewable}\label{reviewing}}

As my analysis has shown, some types of scientific software are
reviewable, but not reviewed today. Several scientific journals
encourage authors to submit code along with their articles, but only a
small number of very specialized journals (e.g.,
\href{https://computo.sfds.asso.fr/}{Computo}, the
\href{https://journalofdigitalhistory.org/}{Journal of Digital History},
\href{https://rescience.github.io/}{ReScience C}) actually review the
submitted code, which tends to be highly situated. Other journals, first
and foremost the \href{https://joss.theoj.org/}{Journal of Open Source
Software}, review software according to generally applicable criteria of
usability and software engineerging practices, but do not expect
reviewers to judge the correctness of the software nor the accuracy or
completeness of its documentation. This would indeed be unrealistic in
the standard journal reviewing process that asks a small number of
individual researchers to evaluate, as volunteers and within short
delays, submissions that are often only roughly in their field of
expertise and represent the work of large teams over many years.

The first category of software that is reviewable but not yet reviewed
is stable wide-spectrum software, such as the GNU Scientific Library
(see section \ref{gsl}). Reviewing could take the form of regular
audits, performed by experts working for an institution dedicated to
this task. In view of the wide use of the software by non-experts in its
domain, the audit should also inspect the software's documentation,
which needs to be up to date and explain the software's functionality
with all the detail that a user must understand. Specifications would be
particularly valuable in this scenario, as the main interface between
developers, users, and auditing experts. For opaque software, formal
specifications could even be made a requirement, in the interest of an
efficient audit. The main difficulty in achieving such audits is that
none of today's scientific institutions consider them part of their
mission.

The second category of reviewable software is transparent software, for
which reviewing consists of mostly the same steps as the established
practices of verification and validation: examining the output for
well-known inputs and/or comparing to the outputs of alternative
implementations of the same tasks. The missing step is the publication
of reports about verification and validation processes performed by
recognized independent experts.

The third category of reviewable software contains situated software,
which can and should be reviewed together with the other outputs of a
research project. For small projects, in terms of the number of
co-authors and the disciplinary spread, situated software could be
reviewed as part of today's peer review process, managed by scientific
journals. The experience of pioneering journals in this activity could
be the basis for elaborating more widely applied reviewing guidelines.
For larger or multidisciplinary projects, the main issue is that today's
peer review process is not adequate at all, even in the (hypothetical)
complete absence of software. Reviewing research performed by a
multidisciplinary team requires another multidisciplinary team, rather
than a few individuals reviewing independently. The integration of
situated software into the process could provide the occasion for a more
general revision of the peer review process.

\hypertarget{science-vs.-the-software-industry}{%
\subsection{Science vs.~the software
industry}\label{science-vs.-the-software-industry}}

In the first decades of computing technology, scientific computing was
one of its main application domains, alongside elaborate bookkeeping
tasks in commerce, finance, and government. Many computers (e.g.~Digital
Equipment Corporation's PDP and VAX series), operating systems
(e.g.~International Business Machine's VM/CMS), and compilers
(e.g.~Fortran) were designed for the needs of scientists. Today,
scientists use mostly computers designed and produced for personal and
business needs. Even supercomputers are constructed to a large degree
from high-grade commodity components that are also found in servers for
cloud computing. Much infrastructure software, such as operating systems
or compilers, are also commodity products developed primarily for other
application domains.

From the perspective of development costs, this evolution makes economic
sense. However, as with any shift towards fewer but more general
products serving a wider client base, the needs of the larger client
groups take priority over those of the smaller ones. Unfortunately for
science, it is today a relative small application domain for software
technology.

In terms of my analysis of reviewability in section \ref{reviewability},
the software industry has a strong focus on proprietary wide-spectrum
software, with a clear distinction between developers and users. Opacity
for users is not seen as a problem, and may even be considered
advantageous as a barrier to reverse-engineering of the software by
competitors. Stability is an expensive characteristic that only
relatively few customers (e.g.~banks or some branches of industry) are
willing to pay for. In contrast, novelty is an important selling
argument in many profitable application domains, leading to attitudes
such as ``move fast and break things'' (the long-time motto of Facebook
founder Mark Zuckerberg), and thus favoring evolving software.

As a consequence of the enormous growth of non-scientific compared to
scientific software, today's dominant software development tools and
software engineering practices largely ignore situated and convivial
software, the impact of dependencies, and the scientific method's
requirement for transparency. However, it can be expected that the
ongoing establishment of Research Software Engineers as a specialization
at the interface between scientific research and software engineering
will lead to development practices that are better aligned with the
specific needs of science. It is such practices that I will propose in
the following sections.

\hypertarget{situated}{%
\subsection{Emphasize situated and convivial software}\label{situated}}

As mentioned in section \ref{wide_spectrum_vs_situated}, many important
scientific software packages are domain-specific tools and libraries,
which have neither the large user base of wide-spectrum software that
justifies external audits, nor the narrow focus of situated software
that allows for a low-effort one-time review by domain experts.
Developing suitable intermediate processes and institutions for
reviewing such software is perhaps possible, but I consider it
scientifically more appropriate to restructure such software into a
convivial collection of more situated modules, supported by a shared
layer that is reviewable due to being wide-spectrum or transparent.
However, this implies assigning a lower priority to reusability, in
conflict with today's best practices in software engineering, and with
recent initiatives to apply the FAIR principles to software {[}Barker et
al. 2022{]}.

In such a scenario, a domain library becomes a collection of source code
files that implement core models and methods, plus ample documentation
of both the methods and implementation techniques. The well-known book
``Numerical Recipes'' {[}Press et al. 2007{]} is a good example for this
approach. Code, documentation, test, and examples need to be written
together, with the explicit goal of supporting users in understanding
and modifying the code, in the spirit of literate programming {[}Knuth
1984{]}.

Users whose needs are not fully met by such a library would make a copy
of the source code files relevant for their work, adapt them to the
particularities of their applications, and make them an integral part of
their own project. In terms of FLOSS jargon, users would make a partial
fork of the project. Version control systems of a kind that doesn't
exist yet would ensure provenance tracking and support the discovery of
other forks. Keeping up to date with relevant forks of one's software
would become part of everyday research work, much like keeping up to
date with publications in one's wider research community. In fact,
another way to describe this approach is full integration of scientific
software development into established research practices, rather than
keeping it a distinct activity governed by different rules. Yet another
perspective is giving priority to the software's role as a
representation of scientific knowledge over its role as a tool {[}Hinsen
2014{]}.

Since this approach differs radically from anything that has been tried
in practice so far, it is premature to discuss its advantages and
downsides. Only practical experience can show to what extent pre-digital
and pre-industrial forms of collaborative knowledge work can be adapted
to automated reasoning. Nevertheless, I will indulge in some speculation
on this topic, to give an idea of what we can fear or hope for.

On the benefit side, the code supporting a specific research project
becomes much smaller and more understandable, mitigating opacity. Its
computational environment is smaller as well, and entirely composed of
stable wide-spectrum or transparent software. Reviewability is therefore
much improved. Moreover, users are encouraged to engage more intensely
with the software, reducing the risk of inappropriate use. The lower
entry barrier to appropriating the code maket it accessible to a wider
range of researchers, increasing inclusiveness and epistemic diversity.

The main loss I expect is in the efficiency of implementing and
deploying new ideas. A strongly coordinated development team whose
members specialize on specific tasks is likely to advance more quickly
in a well-defined direction. This can be a disadvantage in particular
for software whose complexity is dominated by technical rather than
scientific aspects, e.g.~in high-performance computing or large-scale
machine learning applications.

\hypertarget{explainability}{%
\subsection{Make scientific software explainable}\label{explainability}}

Opacity is a major obstacle to the reviewability of software and results
obtained with the help of software, as I have explained in section
\ref{opacity}. Depending on one's precise definition of opacity, it may
be impossible to reduce it. Pragmatically, however, opacity can be
mitigated by \emph{explaining} what the software does, and providing
tools that allow a scientist to \emph{inspect} intermediate or final
results of a computation.

The popularity of computational notebooks, which can be seen as scripts
with attached explanations and results, shows that scientists are indeed
keen on making their work less opaque. But notebooks are limited to the
most situated top layer of a scientific software stack. Code cells in
notebooks refer to library code that can be arbitrarily opaque,
difficult to access, and to which no explanations can be attached.

An interesting line of research in software engineering is exploring
possibilities to make complete software systems explainable
{[}Nierstrasz and Girba 2022; Nierstrasz and Gîrba 2024{]}. The
motivation is similar to explainable AI {[}Gunning et al. 2019{]}, but
the methods are very different, considering that the target of the
explanation is human-authored source code. A new development
environment, \href{https://gtoolkit.com/}{Glamorous Toolkit}
{[}feenk.com 2023{]}, has been designed explicitly to explore these
ideas. Although this work is motivated by situated business
applications, the basic ideas should be transferable to scientific
computing. The screenshots in {[}Nierstrasz and Gîrba 2024{]} convey a
first impression of how these techniques work in practice. However, I
recommend readers to actually download Glamorous Toolkit and explore it
interactively.

The approach is based on three principles. The first one is the same as
for computational notebooks: the integration of code with explanatory
narratives that also contain example code and computed results. Unlike
traditional notebooks, Glamorous Toolkit allows multiple narratives to
reference a shared codebase of arbitrary structure and complexity. The
second principle is the generous use of examples, which serve both as an
illustration for the correct use of the code and as test cases. In
Glamorous Toolkit, whenever you look at some code, you can access
corresponding examples (and also other references to the code) with a
few mouse clicks. The third principle is what the authors call
\emph{moldable inspectors}: situated views on data that present the data
from a domain perspective rather than in terms of its implementation.
These three techniques can be used by software developers to facilitate
the exploration of their systems by others, but they also support the
development process itself by creating new feedback loops.

\hypertarget{specifications}{%
\subsection{Use Digital Scientific Notations}\label{specifications}}

As I have briefly mentioned in the introduction, specifications are
contracts between software developers and software users that describe
the expected behaviour of the software. Formal specifications are
specifications written in a formal language, i.e.~a language amenable to
automated processing. There are various techniques for ensuring or
verifying that a piece of software conforms to a formal specification
{[}Wikipedia 2025a{]}. The use of these tools is, for now, reserved to
software that is critical for safety or security, because of the high
cost of developing specifications and using them to verify
implementations, although the idea of \emph{lightweight} formal methods,
with the potential of much wider applicability, has been circulating for
almost thirty years {[}Jackson and Wing 1996; Zamansky et al. 2018{]}.

Technically, formal specifications are \emph{constraints} on algorithms
and programs, in much the same way as mathematical equations are
constraints on mathematical functions {[}Hinsen 2023{]}. Such
constraints are often much simpler than the algorithms they define. As
an example, consider the task of sorting a list. The (informal)
specification of this task is: produce a new list whose elements are (1)
the same as those of the input list and (2) sorted. A formal version
requires some additional details, in particular a definition of what it
means for two lists to have ``the same'' elements, given that elements
can appear more than once in a list. There are many possible algorithms
conforming to this specification, including well-known sorting
algorithms such as quicksort or bubble sort {[}Wikipedia 2025b{]}. All
of them are much more elaborate than the specification of the result
they produce. As a consequence, they are much more difficult to
understand. Consider an implementation of, say, Quicksort. Many such
implementations are available on
\href{https://rosettacode.org/wiki/Sorting_algorithms/Quicksort}{Rosetta
Code}. It is not obvious from reading such code that it sorts a list. It
is even less obvious that it does so correctly, as there are many
details where a small change would make the results incorrect without
this being obvious. The specification, on the other hand, is immediately
understandable. Moreover, specifications are usually more modular than
algorithms, which also helps human readers to better understand what the
software does {[}Hinsen 2023{]}.

The software engineering contexts in which formal specifications are
used today are very different from the potential applications in
scientific computing that I outline here. In software engineering,
specifications are written to formalize the expected behavior of the
software \emph{before} it is written. The software is considered correct
if it conforms to the specification. In scientific research, software
evolves in parallel with the scientific knowledge that it encodes or
helps to produce. A formal specification has to evolve in the same way,
and is best seen as the formalization of the scientific knowledge.
Change can flow from specification to software, but also in the opposite
direction. Moreoever, most specifications are likely to be incomplete,
leaving out aspects of software behavior that are irrelevant from the
point of view of science (e.g.~resource management or technical
interfaces such as Web APIs), but also aspects that are still under
exploration and thus not yet formalized. For these reasons, I prefer the
term \emph{Digital Scientific Notation} {[}Hinsen 2018b{]}, which better
expresses the role of formal specifications in this context.

Digital Scientific Notations can take many forms. They do not have to
resemble programming languages, nor the specification languages used in
software engineering. My own experimental Digital Scientific Notation,
Leibniz {[}Hinsen 2024a{]}, is intended to resemble traditional
mathematical notation as used e.g.~in physics. Its statements are
embeddable into a narrative, such as a journal article, and it
intentionally lacks typical programming language features such as scopes
that do not exist in natural language, nor in mathematical notation. For
a simple example, see
\href{https://web.archive.org/web/20250130172630/https://leibniz.khinsen.net/lotka-volterra-equations-dx4uzcfvx710f79ajkkgjq1dj.html}{the
Lotka-Volterra equations in Leibniz}. For other domains, graphical
notations may be more appropriate. These notations, the tooling that
integrates them with software, and the scientific practices for working
with them, all remain to be developed. The main expected benefits are
conviviality of the specifications and transparency of the tools that
process them.

\hypertarget{conclusion}{%
\section{Conclusion}\label{conclusion}}

My principal goal with this work is to encourage scientists and research
software engineers to reflect about their computational practices. Why,
and to what degree, do you trust your own computations? What are your
reliability indicators? Do you expose them in your publications? How
reliable does your software have to be to support the conclusions you
draw from their results? Why, and to what degree, do you trust the
computations in the papers you read and cite? Do you consider their
reliability sufficient to support the conclusions made?

These questions are abstract. Answering them requires considering the
concrete level of the specific software used in a computation. The five
categories I have discussed in section \ref{reviewability} should help
with this step, even though it may be difficult at first to evaluate the
software you use on some of the scales. The case studies of section
\ref{casestudies} should help. Situated software is easy to recognize by
its narrow application domain. The size of a software environment is not
difficult to measure, but it requires appropriate tools and training in
their use. The evaluation of stability is often not difficult, but
requires a significant effort, in particular an examination of a
software project's history. Conviviality is hard to diagnose, but rare
anyway. This reduces the examination to Open Source vs.~proprietary,
which is straightforward.

This leaves the transparency vs.~opacity scale, which deserves a more
detailed discussion. Most experienced computational scientists make sure
to examine both intermediate and final results for plausibility, making
use of known properties such as positivity or order of magnitude. But
plausibility is a fuzzy concept. Software is transparent only if users
can check results for \emph{correctness}, not mere plausibility. The
strategies I proposed (sections \ref{situated}, \ref{explainability} and
\ref{specifications}) have the goal of making such correctness checks
easier. If plausibility is all we can check for, then the software is
opaque, and its users are faced with a dilemma when their results are
neither obviously correct nor obviously wrong: are they entitled to
consider them good enough? In practice they do, because the only
realistic alternative would be to stop using computers. Soergel
{[}2015{]} and Thimbleby {[}2023{]} consider this ``trust by default''
misplaced, given what software engineering research tells us about the
frequency of mistakes, and I agree. The examples from the
reproducibility crisis (see the Introduction) support this view that
scientists tend to overestimate the reliability of their work in the
absence of clear signs of problems.

From the perspective of computational reliabilism, the two most critical
characteristics are stability and transparency. If either one is
lacking, it is difficult to consider software to be reliable. Evolving
software changes too frequently to permit the accumulation of evidence
for reliability. Opaque software does not have a clearly defined
behavior. If you don't know what a piece of software is supposed to do
exactly, it is hard to affirm that it does so reliably.

The ideal structure for a reviewable scientific software stack (see Fig.
\ref{fig:software-stack}) whose reliability can actually be argued for
would consist of a foundation of stable infrastructure and a top layer
of situated software (a script, a notebook, or a workflow) that
orchestrates the computations answering a specific scientific question.
The intermediate domain-specific layer would have to be redesigned such
that each of its components is either transparent or small. Digital
scientific notations can help to achieve transparency. Situatedness is a
good strategy for keeping code small.

The main issues we face today are evolving and opaque software, both of
which are common in science. Reaching stability requires time, a large
enough user base, and high software engineering standards. Mitigating
opacity, e.g.~by adopting the strategies I have proposed, requires a
significant effort. Reliability comes at a cost. Making good choices
requires a cost-benefit analysis in the context of each specific
research project. The arguments for the choice should be mentioned in
every research report, to permit readers an assessment of the
reliability of the reported findings.

The difficulty of reviewing scientific software also illustrates the
deficiencies of the current digital infrastructure for
science.\footnote{For more examples, see {[}Saunders 2022{]}.} The
design, implementation, and maintenance of such an infrastructure,
encompassing hardware, software, and best practices, has been neglected
by research institutions, in spite of an overt enthusiasm about the
scientific progress made possible by digital technology. The situation
is improving for research data, for which appropriate repositories and
archives are becoming available. For software, the task is more complex,
and hindered by the contagious neophilia (``tech churn'') of the
software industry. Scientists, research software engineers, research
institutions, and funding agencies must recognize the importance of
stable and reliable infrastructure software, which requires long-term
funding and inclusive governance.

Beyond infrastructural issues, reviewing code is considered impractical
by many researchers because of the enormous effort it seems to require.
However, this argument ignores the fact that we have not yet made a
significant effort to create appropriate reviewing processes. It also
ignores the new possibilities opened up by large language models for
assisting with reviewing code, and in particular reviewing code and its
documentation together as a single entity. But perhaps most importantly,
the argument ignores second-order effects of reviewing. In the long run,
the practice of reviewing scientific code will create an incentive for
scientists and engineers to make their code more reviewable. And the
best time to start the development of reviewing practices is right now.
The use of large language models in software development is growing. It
can help to make software more reviewable, but also less reviewable. The
best way to push the lever to ``more reviewable'' is starting to review
now.

Finally, I need to address a frequent objection to any criticism of the
current use of automated reasoning in science. The objection is that
there is no problem, that science works pretty well overall, because
potential mistakes in an individual scientific study tend to get
detected in a later stage of the eternal knowledge refinement process of
research. While this is true in principle, it is important to understand
the cost of postponing the detection of mistakes and other issues
(biases, dubious approximations, etc.) to later stages. Consider the
case of the five retracted protein structure papers {[}Miller 2006{]}
that I mentioned in the Introduction. They represent a significant
amount of work, much of which much was wasted. In between their
publication and their retraction, other researchers working on similar
proteins had their work rejected because it was in contradiction with
these high-profile publications, meaning more wasted effort. The
undetected software bug caused a lot of frustration, probably ruined the
careers of a few young researchers, and wasted scarce research money. On
the other hand, independent critical inspection of the software would
probably have prevented all of that.\footnote{Assuming that the
  explanation given by the researchers for their mistake is correct. The
  supposedly faulty code has never been made public.}

The question is thus not if we \emph{must} improve computational
practices, but if we \emph{should} do so, in order to reduce frustration
and hardship to researchers, monetary loss to science funders, and lost
opportunities to society at large.

\hypertarget{acknowledgments}{%
\section{Acknowledgments}\label{acknowledgments}}

I would like to thank Nico Formanek {[}Formanek 2025{]} and an anonymous
reviewer {[}Anonymous Reviewer 2025{]} for helpful reviews of an earlier
version of this article, and Juan Durán for an insightful discussion
about computational reliabilism.

\hypertarget{references}{%
\section*{References}\label{references}}
\addcontentsline{toc}{section}{References}

\hypertarget{refs}{}
\begin{CSLReferences}{1}{0}
\leavevmode\vadjust pre{\hypertarget{ref-abrahamGROMACS20253Manual2025}{}}%
\textsc{Abraham, M., Alekseenko, A., Andrews, B., et al.} 2025.
\href{https://doi.org/10.5281/zenodo.16992569}{{GROMACS} 2025.3
{Manual}}.

\leavevmode\vadjust pre{\hypertarget{ref-abrahamGROMACSHighPerformance2015}{}}%
\textsc{Abraham, M.J., Murtola, T., Schulz, R., et al.} 2015.
\href{https://doi.org/10.1016/j.softx.2015.06.001}{{GROMACS}: {High}
performance molecular simulations through multi-level parallelism from
laptops to supercomputers}. \emph{SoftwareX} \emph{1--2}, 19--25.

\leavevmode\vadjust pre{\hypertarget{ref-anonymousreviewerReviewEstablishingTrust2025}{}}%
\textsc{Anonymous Reviewer}. 2025.
\href{https://doi.org/10.70744/MetaROR.39.1.rv1}{Review: {Establishing}
trust in automated reasoning ({Round} 1 - {Review} 1)}.

\leavevmode\vadjust pre{\hypertarget{ref-baker1500ScientistsLift2016}{}}%
\textsc{Baker, M.} 2016. \href{https://doi.org/10.1038/533452a}{1,500
scientists lift the lid on reproducibility}. \emph{Nature} \emph{533},
7604, 452--454.

\leavevmode\vadjust pre{\hypertarget{ref-barkerIntroducingFAIRPrinciples2022}{}}%
\textsc{Barker, M., Chue Hong, N.P., Katz, D.S., et al.} 2022.
\href{https://doi.org/10.1038/s41597-022-01710-x}{Introducing the {FAIR
Principles} for research software}. \emph{Scientific Data} \emph{9}, 1,
622.

\leavevmode\vadjust pre{\hypertarget{ref-bilkeReproducibleHPCSoftware2025}{}}%
\textsc{Bilke, L., Fischer, T., Naumov, D., and Meisel, T.} 2025.
\href{https://doi.org/10.1007/s12665-025-12501-z}{Reproducible {HPC}
software deployments, simulations, and workflows -- a case study for
far-field deep geological repository assessment}. \emph{Environmental
Earth Sciences} \emph{84}, 17, 502.

\leavevmode\vadjust pre{\hypertarget{ref-williamsChapterWantPrinter2002}{}}%
\textsc{Chapter 1: {For Want} of a {Printer}}. 2002. In: \emph{Free as
in freedom: {Richard Stallman}'s crusade for free software}. O'Reilly,
Sebastopol, Calif. : Farnham.

\leavevmode\vadjust pre{\hypertarget{ref-colliardEconomicsComputationalReproducibility2022}{}}%
\textsc{Colliard, J.-E., Hurlin, C., and Pérignon, C.} 2022.
\href{https://doi.org/10.2139/ssrn.3418896}{The {Economics} of
{Computational Reproducibility}}.

\leavevmode\vadjust pre{\hypertarget{ref-courtesWhatPackage2021}{}}%
\textsc{Courtès, L.} 2021. What's in a package. \emph{GuixHPC blog}.

\leavevmode\vadjust pre{\hypertarget{ref-duboisNumericalPython1996}{}}%
\textsc{Dubois, P.F., Hinsen, K., and Hugunin, J.} 1996.
\href{https://doi.org/10.1063/1.4822400}{Numerical {Python}}.
\emph{Computers in Physics} \emph{10}, 3, 262.

\leavevmode\vadjust pre{\hypertarget{ref-duranTransparencyComputationalReliabilism2025}{}}%
\textsc{Durán, J.M.} 2025. Beyond transparency: Computational
reliabilism as an externalist epistemology of algorithms.
\url{https://arxiv.org/abs/2502.20402}.

\leavevmode\vadjust pre{\hypertarget{ref-duranGroundsTrustEssential2018}{}}%
\textsc{Durán, J.M. and Formanek, N.} 2018.
\href{https://doi.org/10.1007/s11023-018-9481-6}{Grounds for {Trust}:
{Essential Epistemic Opacity} and {Computational Reliabilism}}.
\emph{Minds and Machines} \emph{28}, 4, 645--666.

\leavevmode\vadjust pre{\hypertarget{ref-feenk.comGlamorousToolkit2023}{}}%
\textsc{feenk.com}. 2023. Glamorous {Toolkit}.

\leavevmode\vadjust pre{\hypertarget{ref-formanekReviewEstablishingTrust2025}{}}%
\textsc{Formanek, N.} 2025.
\href{https://doi.org/10.70744/MetaROR.39.1.rv2}{Review: {Establishing}
trust in automated reasoning ({Round} 1 - {Review} 2)}.

\leavevmode\vadjust pre{\hypertarget{ref-frenchministryforhighereducationresearchandinnovationSecond_French_PlanforOpenScience_web2021}{}}%
\textsc{French Ministry for Higher Education, Research, and Innovation}.
2021. \emph{Second\_{French}\_{Plan-for-Open-Science}\_web}.

\leavevmode\vadjust pre{\hypertarget{ref-galassiGNUScientificLibrary2009}{}}%
\textsc{Galassi, M., ed.} 2009. \emph{{GNU} scientific library reference
manual: For {GSL} version 1.12}. Network Theory, Bristol.

\leavevmode\vadjust pre{\hypertarget{ref-georgiouExperimentersSexModulates2022}{}}%
\textsc{Georgiou, P., Zanos, P., Mou, T.-C.M., et al.} 2022.
\href{https://doi.org/10.1038/s41593-022-01146-x}{Experimenters' sex
modulates mouse behaviors and neural responses to ketamine via
corticotropin releasing factor}. \emph{Nature Neuroscience} \emph{25},
9, 1191--1200.

\leavevmode\vadjust pre{\hypertarget{ref-gonzalez-barahonaBriefHistoryFree2021}{}}%
\textsc{Gonzalez-Barahona, J.M.} 2021.
\href{https://doi.org/10.1109/MC.2020.3041887}{A {Brief History} of
{Free}, {Open Source Software} and {Its Communities}}. \emph{Computer}
\emph{54}, 2, 75--79.

\leavevmode\vadjust pre{\hypertarget{ref-GSLGNUScientific2025}{}}%
\textsc{{GSL} - {GNU Scientific Library} - {GNU Project} - {Free
Software Foundation}}. 2025.

\leavevmode\vadjust pre{\hypertarget{ref-gunningXAIExplainableArtificial2019}{}}%
\textsc{Gunning, D., Stefik, M., Choi, J., Miller, T., Stumpf, S., and
Yang, G.-Z.} 2019.
\href{https://doi.org/10.1126/scirobotics.aay7120}{{XAI}---{Explainable}
artificial intelligence}. \emph{Science Robotics} \emph{4}, 37,
eaay7120.

\leavevmode\vadjust pre{\hypertarget{ref-harrisArrayProgrammingNumPy2020a}{}}%
\textsc{Harris, C.R., Millman, K.J., van der Walt, S.J., et al.} 2020.
\href{https://doi.org/10.1038/s41586-020-2649-2}{Array programming with
{NumPy}}. \emph{Nature} \emph{585}, 7825, 357--362.

\leavevmode\vadjust pre{\hypertarget{ref-hasselbringMultidimensionalResearchSoftware2025}{}}%
\textsc{Hasselbring, W., Druskat, S., Bernoth, J., et al.} 2025.
\href{https://doi.org/10.1109/MCSE.2025.3555023}{Multidimensional
{Research Software Categorization}}. \emph{Computing in Science \&
Engineering} \emph{27}, 2, 59--68.

\leavevmode\vadjust pre{\hypertarget{ref-herndonDoesHighPublic2014}{}}%
\textsc{Herndon, T., Ash, M., and Pollin, R.} 2014.
\href{https://doi.org/10.1093/cje/bet075}{Does high public debt
consistently stifle economic growth? {A} critique of {Reinhart} and
{Rogoff}}. \emph{Cambridge Journal of Economics} \emph{38}, 2, 257--279.

\leavevmode\vadjust pre{\hypertarget{ref-hettrickResearchSoftwareSustainability2016}{}}%
\textsc{Hettrick, S.} 2016. \emph{Research {Software Sustainability}}.
Knowledge Exchange.

\leavevmode\vadjust pre{\hypertarget{ref-hettrickUKResearchSoftware2014}{}}%
\textsc{Hettrick, S., Antonioletti, M., Carr, L., et al.} 2014.
\href{https://doi.org/10.5281/zenodo.14809}{{UK Research Software
Survey} 2014}.

\leavevmode\vadjust pre{\hypertarget{ref-hinsenComputationalScienceShifting2014}{}}%
\textsc{Hinsen, K.} 2014.
\href{https://doi.org/10.12688/f1000research.3978.2}{Computational
science: Shifting the focus from tools to models}. \emph{F1000Research}
\emph{3}, 101.

\leavevmode\vadjust pre{\hypertarget{ref-hinsenPowerCreateChaos2016}{}}%
\textsc{Hinsen, K.} 2016.
\href{https://doi.org/10.1109/MCSE.2016.67}{The {Power} to {Create
Chaos}}. \emph{Computing in Science \& Engineering} \emph{18}, 4,
75--79.

\leavevmode\vadjust pre{\hypertarget{ref-hinsenReusableReeditableCode2018}{}}%
\textsc{Hinsen, K.} 2018a.
\href{https://doi.org/10.1109/MCSE.2018.03202636}{Reusable {Versus}
{Re-editable Code}}. \emph{Computing in Science \& Engineering}
\emph{20}, 3, 78--83.

\leavevmode\vadjust pre{\hypertarget{ref-hinsenVerifiabilityComputeraidedResearch2018}{}}%
\textsc{Hinsen, K.} 2018b.
\href{https://doi.org/10.7717/peerj-cs.158}{Verifiability in
computer-aided research: The role of digital scientific notations at the
human-computer interface}. \emph{PeerJ Computer Science} \emph{4}, e158.

\leavevmode\vadjust pre{\hypertarget{ref-hinsenDealingSoftwareCollapse2019}{}}%
\textsc{Hinsen, K.} 2019.
\href{https://doi.org/10.1109/MCSE.2019.2900945}{Dealing {With Software
Collapse}}. \emph{Computing in Science \& Engineering} \emph{21}, 3,
104--108.

\leavevmode\vadjust pre{\hypertarget{ref-hinsenComputationalReproducibility2020}{}}%
\textsc{Hinsen, K.} 2020. Computational {Reproducibility}. In:
\emph{Computation in {Science}: {From} concepts to practice}. IOP
Publishing.

\leavevmode\vadjust pre{\hypertarget{ref-hinsenNatureComputationalModels2023}{}}%
\textsc{Hinsen, K.} 2023.
\href{https://doi.org/10.1109/MCSE.2023.3286250}{The nature of
computational models}. \emph{Computing In Science \& Engineering}
\emph{25}, 1, 61--66.

\leavevmode\vadjust pre{\hypertarget{ref-hinsenLeibnizDigitalScientific2024}{}}%
\textsc{Hinsen, K.} 2024a. Leibniz - a {Digital Scientific Notation}.

\leavevmode\vadjust pre{\hypertarget{ref-hinsenRedressingBalanceYinYang2024}{}}%
\textsc{Hinsen, K.} 2024b.
\href{https://doi.org/10.1145/3689492.3689808}{Redressing the {Balance}:
{A Yin-Yang Perspective} on {Information Technology}}. \emph{Proceedings
of the 2024 {ACM SIGPLAN International Symposium} on {New Ideas}, {New
Paradigms}, and {Reflections} on {Programming} and {Software}},
Association for Computing Machinery, 194--204.

\leavevmode\vadjust pre{\hypertarget{ref-hinsenReviewingResearchSoftware2025}{}}%
\textsc{Hinsen, K.} 2025.
\href{https://doi.org/10.1109/MCSE.2025.3601909}{Reviewing {Research
Software}}. \emph{Computing in Science \& Engineering} \emph{27}, 3,
64--66.

\leavevmode\vadjust pre{\hypertarget{ref-hocquetOnlyInitiatesWill2017}{}}%
\textsc{Hocquet, A. and Wieber, F.} 2017.
\href{https://doi.org/10.1109/MAHC.2018.1221048}{{``{Only} the
{Initiates Will Have} the {Secrets Revealed}''}: {Computational
Chemists} and the {Openness} of {Scientific Software}}. \emph{IEEE
Annals of the History of Computing} \emph{39}, 4, 40--58.

\leavevmode\vadjust pre{\hypertarget{ref-humphreysPhilosophicalNoveltyComputer2009}{}}%
\textsc{Humphreys, P.} 2009.
\href{https://doi.org/10.1007/s11229-008-9435-2}{The {Philosophical
Novelty} of {Computer Simulation Methods}}. \emph{Synthese} \emph{169},
3, 615--626.

\leavevmode\vadjust pre{\hypertarget{ref-illichToolsConviviality1973}{}}%
\textsc{Illich, I.} 1973. \emph{Tools for conviviality}. {Calders and
Boyars}, London.

\leavevmode\vadjust pre{\hypertarget{ref-jacksonLightweightFormalMethods1996}{}}%
\textsc{Jackson, D. and Wing, J.M.} 1996. Lightweight {Formal Methods}.
\emph{IEEE Computer} \emph{April 1996}, 21--22.

\leavevmode\vadjust pre{\hypertarget{ref-jumperHighlyAccurateProtein2021}{}}%
\textsc{Jumper, J., Evans, R., Pritzel, A., et al.} 2021.
\href{https://doi.org/10.1038/s41586-021-03819-2}{Highly accurate
protein structure prediction with {AlphaFold}}. \emph{Nature}
\emph{596}, 7873, 583--589.

\leavevmode\vadjust pre{\hypertarget{ref-kayPersonalComputerChildren1972}{}}%
\textsc{Kay, A.C.} 1972. \href{https://doi.org/10.1145/800193.1971922}{A
{Personal Computer} for {Children} of {All Ages}}. \emph{Proceedings of
the {ACM} annual conference - {Volume} 1}, Association for Computing
Machinery.

\leavevmode\vadjust pre{\hypertarget{ref-kellConvivialDesignHeuristics2020}{}}%
\textsc{Kell, S.} 2020.
\href{https://doi.org/10.1145/3397537.3397543}{Convivial design
heuristics for software systems}. \emph{Conference {Companion} of the
4th {International Conference} on {Art}, {Science}, and {Engineering} of
{Programming}}, ACM, 144--148.

\leavevmode\vadjust pre{\hypertarget{ref-knuthLiterateProgramming1984}{}}%
\textsc{Knuth, D.E.} 1984. Literate programming. \emph{The Computer
Journal} \emph{27}, 2, 97--111.

\leavevmode\vadjust pre{\hypertarget{ref-koStateArtEnduser2011}{}}%
\textsc{Ko, A.J., Abraham, R., Beckwith, L., et al.} 2011.
\href{https://doi.org/10.1145/1922649.1922658}{The state of the art in
end-user software engineering}. \emph{ACM Computing Surveys} \emph{43},
3, 21:1--21:44.

\leavevmode\vadjust pre{\hypertarget{ref-kortzfleischMultipleExperimentersImprove2022}{}}%
\textsc{Kortzfleisch, V.T. von, Ambrée, O., Karp, N.A., et al.} 2022.
\href{https://doi.org/10.1371/journal.pbio.3001564}{Do multiple
experimenters improve the reproducibility of animal studies?} \emph{PLOS
Biology} \emph{20}, 5, e3001564.

\leavevmode\vadjust pre{\hypertarget{ref-leonelliOpenScienceEpistemic2022}{}}%
\textsc{Leonelli, S.} 2022.
\href{https://doi.org/10.1017/psa.2022.45}{Open {Science} and {Epistemic
Diversity}: {Friends} or {Foes}?} \emph{Philosophy of Science}
\emph{89}, 5, 991--1001.

\leavevmode\vadjust pre{\hypertarget{ref-longoExplainableArtificialIntelligence2024}{}}%
\textsc{Longo, L., Brcic, M., Cabitza, F., et al.} 2024.
\href{https://doi.org/10.1016/j.inffus.2024.102301}{Explainable
{Artificial Intelligence} ({XAI}) 2.0: {A} manifesto of open challenges
and interdisciplinary research directions}. \emph{Information Fusion}
\emph{106}, 102301.

\leavevmode\vadjust pre{\hypertarget{ref-malkaReproducibilityBuildEnvironments2024}{}}%
\textsc{Malka, J., Zacchiroli, S., and Zimmermann, T.} 2024.
Reproducibility of {Build Environments} through {Space} and {Time}.
\emph{46th {International Conference} on {Software Engineering} ({ICSE}
2024) - {New Ideas} and {Emerging Results} ({NIER}) {Track}}.

\leavevmode\vadjust pre{\hypertarget{ref-meraliComputationalScienceError2010}{}}%
\textsc{Merali, Z.} 2010.
\href{https://doi.org/10.1038/467775a}{Computational science:
...{Error}}. \emph{Nature} \emph{467}, 7317, 775--777.

\leavevmode\vadjust pre{\hypertarget{ref-millerScientistsNightmareSoftware2006}{}}%
\textsc{Miller, G.} 2006.
\href{https://doi.org/10.1126/science.314.5807.1856}{A {Scientist}'s
{Nightmare}: {Software Problem Leads} to {Five Retractions}}.
\emph{Science} \emph{314}, 5807, 1856--1857.

\leavevmode\vadjust pre{\hypertarget{ref-nangiaUnderstandingSoftwareResearch2017}{}}%
\textsc{Nangia, U. and Katz, D.S.} 2017.
\href{https://doi.org/10.1109/eScience.2017.78}{Understanding {Software}
in {Research}: {Initial Results} from {Examining Nature} and a {Call}
for {Collaboration}}. \emph{2017 {IEEE} 13th {International Conference}
on e-{Science} (e-{Science})}, 486--487.

\leavevmode\vadjust pre{\hypertarget{ref-nardiSmallMatterProgramming1993}{}}%
\textsc{Nardi, B.A.} 1993. \emph{A small matter of programming:
Perspectives on end user computing}. MIT Press, Cambridge, MA.

\leavevmode\vadjust pre{\hypertarget{ref-nielsenHowAIImpacting2023}{}}%
\textsc{Nielsen, M.} 2023. How is {AI} impacting science?

\leavevmode\vadjust pre{\hypertarget{ref-nierstraszMakingSystemsExplainable2022}{}}%
\textsc{Nierstrasz, O. and Girba, T.} 2022.
\href{https://doi.org/10.1109/VISSOFT55257.2022.00009}{Making {Systems
Explainable}}. \emph{2022 {Working Conference} on {Software
Visualization} ({VISSOFT})}, IEEE, 1--4.

\leavevmode\vadjust pre{\hypertarget{ref-nierstraszMoldableDevelopmentPatterns2024}{}}%
\textsc{Nierstrasz, O.M. and Gîrba, T.} 2024.
\href{https://doi.org/10.1145/3698322.3698327}{Moldable {Development
Patterns}}. \emph{Proceedings of the 29th {European Conference} on
{Pattern Languages} of {Programs}, {People}, and {Practices}},
Association for Computing Machinery, 1--14.

\leavevmode\vadjust pre{\hypertarget{ref-parasuramanHumansAutomationUse1997}{}}%
\textsc{Parasuraman, R. and Riley, V.} 1997.
\href{https://doi.org/10.1518/001872097778543886}{Humans and
{Automation}: {Use}, {Misuse}, {Disuse}, {Abuse}}. \emph{Human Factors}
\emph{39}, 2, 230--253.

\leavevmode\vadjust pre{\hypertarget{ref-poteStabilityDesign2025}{}}%
\textsc{Pote, T.} 2025. Stability by {Design}.

\leavevmode\vadjust pre{\hypertarget{ref-pressNumericalRecipesArt2007}{}}%
\textsc{Press, W.H., Teukolsky, S.A., Vetterling, W.T., and Flannery,
B.P.} 2007. \emph{Numerical recipes: The art of scientific computing}.
Cambridge University Press, Cambridge, UK ; New York.

\leavevmode\vadjust pre{\hypertarget{ref-samuelComputationalReproducibilityJupyter2024}{}}%
\textsc{Samuel, S. and Mietchen, D.} 2024.
\href{https://doi.org/10.1093/gigascience/giad113}{Computational
reproducibility of {Jupyter} notebooks from biomedical publications}.
\emph{GigaScience} \emph{13}, giad113.

\leavevmode\vadjust pre{\hypertarget{ref-saundersDecentralizedInfrastructureNeuro2022}{}}%
\textsc{Saunders, J.L.} 2022. Decentralized {Infrastructure} for
({Neuro})science. \url{https://arxiv.org/abs/2209.07493}.

\leavevmode\vadjust pre{\hypertarget{ref-shawMythsMythconceptionsWhat2022}{}}%
\textsc{Shaw, M.} 2022. \href{https://doi.org/10.1145/3480947}{Myths and
mythconceptions: What does it mean to be a programming language,
anyhow?} \emph{Proceedings of the ACM on Programming Languages}
\emph{4}, HOPL, 1--44.

\leavevmode\vadjust pre{\hypertarget{ref-shirkySituatedSoftware2004}{}}%
\textsc{Shirky, C.} 2004. Situated {Software}.

\leavevmode\vadjust pre{\hypertarget{ref-soergelRampantSoftwareErrors2015}{}}%
\textsc{Soergel, D.A.W.} 2015.
\href{https://doi.org/10.12688/f1000research.5930.2}{Rampant software
errors may undermine scientific results}.

\leavevmode\vadjust pre{\hypertarget{ref-spierHistoryPeerreviewProcess2002}{}}%
\textsc{Spier, R.} 2002.
\href{https://doi.org/10.1016/S0167-7799(02)01985-6}{The history of the
peer-review process}. \emph{Trends in Biotechnology} \emph{20}, 8,
357--358.

\leavevmode\vadjust pre{\hypertarget{ref-stoddenBestPracticesComputational2013}{}}%
\textsc{Stodden, V. and Miguez, S.} 2013.
\href{https://doi.org/10.5334/jors.ay}{Best {Practices} for
{Computational Science}: {Software Infrastructure} and {Environments}
for {Reproducible} and {Extensible Research}}. \emph{Social Science
Research Network} \emph{2322276}.

\leavevmode\vadjust pre{\hypertarget{ref-stoddenEmpiricalAnalysisJournal2018a}{}}%
\textsc{Stodden, V., Seiler, J., and Ma, Z.} 2018.
\href{https://doi.org/10.1073/pnas.1708290115}{An empirical analysis of
journal policy effectiveness for computational reproducibility}.
\emph{Proceedings of the National Academy of Sciences} \emph{115}, 11,
2584--2589.

\leavevmode\vadjust pre{\hypertarget{ref-thimblebyImprovingScienceThat2023}{}}%
\textsc{Thimbleby, H.} 2023.
\href{https://doi.org/10.1093/comjnl/bxad067}{Improving {Science That
Uses Code}}. \emph{The Computer Journal}, bxad067.

\leavevmode\vadjust pre{\hypertarget{ref-thompsonReflectionsTrustingTrust1984}{}}%
\textsc{Thompson, K.} 1984.
\href{https://doi.org/10.1145/358198.358210}{Reflections on trusting
trust}. \emph{Communications of the ACM} \emph{27}, 8, 761--763.

\leavevmode\vadjust pre{\hypertarget{ref-turkleSimulationItsDiscontents2009}{}}%
\textsc{Turkle, S.} 2009. \emph{Simulation and its discontents}. The MIT
Press, Cambridge, Massachusetts.

\leavevmode\vadjust pre{\hypertarget{ref-wangRestoringExecutionEnvironments2021}{}}%
\textsc{Wang, J., Li, L., and Zeller, A.} 2021.
\href{https://doi.org/10.1109/ICSE43902.2021.00144}{Restoring {Execution
Environments} of {Jupyter Notebooks}}. \emph{2021 {IEEE}/{ACM} 43rd
{International Conference} on {Software Engineering} ({ICSE})},
1622--1633.

\leavevmode\vadjust pre{\hypertarget{ref-wikipediaFormalMethods2025}{}}%
\textsc{Wikipedia}. 2025a. Formal methods. \emph{Wikipedia}.

\leavevmode\vadjust pre{\hypertarget{ref-wikipediaSortingAlgorithm2025}{}}%
\textsc{Wikipedia}. 2025b. Sorting algorithm. \emph{Wikipedia}.

\leavevmode\vadjust pre{\hypertarget{ref-wilkinsonFAIRGuidingPrinciples2016}{}}%
\textsc{Wilkinson, M.D., Dumontier, M., Aalbersberg, Ij.J., et al.}
2016. \href{https://doi.org/10.1038/sdata.2016.18}{The {FAIR Guiding
Principles} for scientific data management and stewardship}.
\emph{Scientific Data} \emph{3}, 160018.

\leavevmode\vadjust pre{\hypertarget{ref-zamanskyClassificationLightweightFormal2018}{}}%
\textsc{Zamansky, A., Spichkova, M., Rodriguez-Navas, G., Herrmann, P.,
and Blech, J.O.} 2018.
\href{https://doi.org/10.5220/0006770803050313}{Towards {Classification}
of {Lightweight Formal Methods}}. \emph{Proceedings of the 13th
{International Conference} on {Evaluation} of {Novel Approaches} to
{Software Engineering}}, {SCITEPRESS - Science and Technology
Publications, Lda}, 305--313.

\end{CSLReferences}

\end{document}